\documentclass[aip,jcp,amsmath,amssymb,superscriptaddress,12pt,reprint]{revtex4-1}

\usepackage{graphicx}
\usepackage{bm}
\usepackage{microtype}
\usepackage[breaklinks=true,hyperindex=true,
colorlinks=true,pagebackref=false,citecolor=blue,plainpages=false,pdfpagelabels,
linkcolor=blue,urlcolor=blue]{hyperref}
\usepackage{xcolor}
\newcommand{\sigx}{\hat \varrho}
\newcommand{\densityR}{{\boldsymbol R}}

\newcommand{\eigenmarker}{}

\def\proj#1#2{\left| #1 \left\rangle \right\langle #2\right|}

\newcommand{\CPTG}{
Chemical Physics Theory Group, Department of Chemistry, University of Toronto, Toronto, Ontario M5S 3H6, Canada.}

\newcommand{\CQIQC}{
Center for Quantum Information and Quantum Control, University of Toronto, Toronto, Ontario M5S 3H6, Canada}

\newcommand{\GFTyMAyGFAM}{
Grupo de F{\'i}sica Te{\'o}rica y Matem{\'a}tica Aplicada; y Grupo de F\'isica At\'omica y Molecular; Instituto de F{\'i}sica,
Facultad de Ciencias Exactas y Naturales,
Universidad de Antioquia; Calle 70 No.~52-21, Medell\'in, Colombia.}

\begin{document}

\title{Equilibrium Coherence in the Multi-level Spin-boson Model}

\author{Mike Reppert}
\email[]{mreppert@alum.mit.edu}
\affiliation{\CPTG}

\author{Deborah Reppert}
\affiliation{\CPTG}

\author{Leonardo A. Pachon}
\affiliation{\GFTyMAyGFAM}

\author{Paul Brumer}
\affiliation{\CPTG}
\affiliation{\CQIQC}

\date{\today}

\begin{abstract}
Interaction between a quantum system and its environment can induce stationary
coherences -- off-diagonal elements in the reduced system density matrix -- even
at equilibrium.
This work investigates the ``quantumness'' of such phenomena by examining the ability of classical and semiclassical models to describe equilibrium stationary coherence in the multi-level spin boson (MLSB) model, a common model for light-harvesting systems.
A well justified classical harmonic oscillator model is found to fail to capture equilibrium coherence. This failure is attributed to the effective weakness of classical system-bath interactions due to the absence of a discrete system energy spectrum and, consequently, of quantized shifts in oscillator coordinates.
Semiclassical coherences also vanish for a dimeric model with parameters typical of biological light-harvesting, i.e., where both system sites couple to the bath with the same reorganization energy.
In contrast, equilibrium coherence persists in a fully quantum description of the same system, suggesting a uniquely quantum-mechanical origin for equilibrium stationary coherence in, e.g., photosynthetic systems.
Finally, as a computational tool,  a perturbative expansion is introduced that, at third order in $\hbar$, gives qualitatively correct behavior at ambient temperatures for all
configurations examined.
\end{abstract}


\maketitle 

\newcommand{\ket}[1]{\left| #1 \right>} 
\newcommand{\bra}[1]{\left< #1 \right|} 
\newcommand{\braket}[2]{\left< #1 \vphantom{#2} \right|
\left. #2 \vphantom{#1} \right>} 
\newcommand{\braAket}[3]{\left< #1 \left| #2
\right| #3 \right>} 

\newcommand{\lket}[1]{\left| #1 \right)} 
\newcommand{\lbra}[1]{\left( #1 \right|} 
\newcommand{\lbraket}[2]{\left( #1 \vphantom{#2} \right|
 \left. #2 \vphantom{#1} \right)} 
\newcommand{\lbraAket}[3]{\left( #1 \left| #2
\right| #3 \right)} 
\linespread{1.0}

\section{Introduction}
System-bath interactions are typically viewed as deleterious to quantum coherence within the system. Yet in some cases, interactions with the environment can  {\it induce} coherence in an open quantum system even under stationary conditions where the density matrix would otherwise be expected to be diagonal.\cite{Moix2012,Lee2012,PB13a,PB13b,PYB13,Dodin2016,Ma2015,Tscherbul2015,Tscherbul2018,Guarnieri2018,PT&19,dodin2019}
Such deviations from canonical system statistics often persist even under physiological conditions and can have a significant effect on biological processes such as visual phototransduction and photosynthetic light harvesting.\cite{Tscherbul2015,Ma2015,dodin2019}
These environment-induced stationary coherences -- off-diagonal elements in the reduced system density matrix in the eigenbasis of the system Hamiltonian -- have attracted particular attention recently as a potential resource for the enhanced performance of quantum devices.\cite{Scully2011,PB13a,PB13b,Holubec2018,Dorfman2018} Despite these developments, little progress has been made in characterizing the conditions under which stationary coherence represents an intrinsically quantum-mechanical phenomenon \cite{PT&19}.
This paper explores this question by investigating the extent to which classical and semiclassical models exhibit stationary coherence at equilibrium, focusing on the multi-level spin-boson (MLSB) model relevant for photosynthetic light-harvesting dynamics.

The question of the ``quantumness'' of equilibrium coherence is particularly timely in light of recent observations that many {\it dynamic} coherence effects in quantum systems can be reproduced by classical models.\cite{Eisfeld2012,Briggs2012,Briggs2011,Briggs2013,Mancal2013,Reppert2018,Reppert2018b} These recent findings are closely connected to the classic demonstrations by Meyer, Miller, Stock, and Thoss of strong parallels between the classical dynamics of coupled harmonic oscillators and the wavefunction dynamics of finite quantum systems, even in the presence of a dephasing environment.\cite{Meyer1979,Stock1997,Stock2005} The motivation for the present study is that, while such classical models have been shown to provide accurate descriptions of short-time {\it dynamic} coherence effects, it is unclear whether they are capable of describing long-time {\it equilibrium} coherence effects.

The organization of this paper is as follows. Section~\ref{sec_model} introduces the quantum MLSB model and the phenomenon of equilibrium coherence.
Section~\ref{sec_classical} describes a classical analog of the MLSB model and shows that it fails to exhibit stationary coherence at equilibrium.
Section~\ref{sec_semiclassical} introduces a semiclassical framework for stationary coherence.
Section~\ref{sec_quantum} quantitatively compares these semiclassical coherences with their fully-quantum analogs at second order in the system-bath interaction strength and identifies conditions under which classical and semiclassical models fail to capture quantum coherence effects in the MLSB model.
Section~\ref{sec_hbar_expansion} introduces a small-$\hbar$ expansion for interpolating between quantum and classical limits, and Section~\ref{sec_phase_space} discusses the physical origins of coherence in these various descriptions using a phase-space representation.
Section~\ref{sec_quantumness} discusses the implications of the findings for quantum effects in light-harvesting devices, and Section~\ref{sec_conclusions} summarizes the findings.

\section{Equilibrium Coherence in the MLSB}
\label{sec_model}
Light-harvesting in photosynthetic pigment-protein complexes (PPCs) is often theoretically described as a system of Frenkel excitons coupled to a harmonic bath, termed the multi-level spin boson (MLSB) model.
The material Hamiltonian reads\cite{Renger2002,Cheng2009,Ishizaki2009,Zimanyi2012,Tiwari2013}
\begin {eqnarray}
	\label{def_H}
	\hat H = \hat{H}_\mathrm{l} + \hat{H}_\mathrm{ss} + \hat{H}_\mathrm{SB} + \hat{H}_\mathrm{B}
\end {eqnarray}
where
\begin{eqnarray}
\label{Hl_quantum}
\hat{H}_\mathrm{l} &=& \sum_n \hbar \omega_n \ket{n}\bra{n},
 \\
	\label{Hss_quantum}
	\hat{H}_\mathrm{ss} &=& \sum_{n,m} \hbar V_{nm} \ket{n}\bra{m} ,
\\
	\label{HSB_quantum}
	\hat{H}_\mathrm{SB} &=& \sum_{n,k} \hbar \alpha_{nk} \hat Q_k \ket{n}\bra{n} ,
\\
	\label{HB_quantum}
	\hat{H}_\mathrm{B} &=& \frac{1}{2} \sum_k \left( \Omega_k^2 \hat Q_k^2 + \hat P_k^2 \right).
\end{eqnarray}
Here the states $\{\ket {n} \}$ represent the lowest electronic excited state of the $n^\text{th}$ pigment in an $N_\mathrm{S}$-site pigment-protein complex, with all other pigments in their electronic ground state. The total electronic ground state $\ket{0}$, in which no pigments are excited, is taken to have energy zero and so is included only implicitly in the Hamiltonian. Throughout this work, electronic summation indices (e.g., $m$, and $n$) are understood to run over the values $1, ..., N_\mathrm{S}$, not including the ground state, unless otherwise noted.
These local excitations interact with one another via the site-to-site interaction Hamiltonian $\hat{H}_\mathrm{ss}$ and with a harmonic bath (Hamiltonian $\hat{H}_\mathrm{B}$) via the system-bath interaction Hamiltonian $\hat{H}_\mathrm{SB}$. The relative strengths of system-system and system-bath interactions are determined by the magnitude of the coupling coefficients $V_{nm}$ and $\alpha_{nk}$.
%
Model parameters vary between different PPCs but are subject to the restrictions
\begin{eqnarray}
	\label{quantum_adiabatic_restriction}
	\bar\omega \gg \omega_{mn}, V_{mn}, \Omega_k,  \frac{E_{mn}^\mathrm{r}}{\hbar}
\end{eqnarray}
and
\begin {eqnarray}
	\label {quantum_temperature_restriction}
	\hbar\bar\omega\gg k_\mathrm{B} T,
\end {eqnarray}
where $k_\mathrm{B} T$ is the thermodynamic temperature, $\bar\omega =\frac {1} {N_\mathrm{S}} \sum_n \omega_n $ is the average frequency of the local sites, $\omega_{m n} = \omega_m - \omega_n $, and
\begin {eqnarray}
	\label {lambda_definition}
	E_{mn}^\mathrm{r} \equiv \hbar^2 \sum_k \frac {\alpha_{mk} \alpha_{nk} } {2 \Omega_k^2 }
\end {eqnarray}
is the reorganization energy.
These restrictions ensure that the electronic ground and excited states do not mix and that the thermal population of electronic excited states is negligible.

The off-diagonal reorganization energies $E_{mn}^\mathrm{r}$ indicate the degree of correlation between system-bath interactions at each site. Three cases are particularly noteworthy. In the case of \emph{perfectly correlated} system-bath interactions, the coupling coefficents $\alpha_{nk}$ are independent of $n$ (the system site). In this case, all system frequencies fluctuate in sync with each other when viewed as parametric functions of the bath coordinates $\hat Q_k$, and the reorganization energy $E_{mn}^\text{r}$ is independent of both $m$ and $n$. In contrast, perfect \emph{anticorrelation} (for a dimer) implies that $\alpha_{1k} = - \alpha_{2k}$ for all $k$, so that $E_{12}^\mathrm{r} =E_{21}^\mathrm{r} = - E_{11}^\mathrm{r} = - E_{22}^\mathrm{r}$. In this case, bath-dependent frequency shifts are equal in magnitude but opposite in sign at the two sites. As has recently been emphasized,\cite{Tiwari2013} anticorrelated bath modes play a central role in the energy transfer dynamics of the MLSB model. Finally, \emph{uncorrelated} system-bath interactions occur when each bath mode $Q_k$ couples only to a single pigment, so that $\alpha_{mk}\alpha_{nk} = 0$ for $m \neq n$; in this case, $E_{mn}^\mathrm{r} = 0$ for $m\neq n$, and there is no correlation between bath-dependent frequency shifts at each site.

This work investigates stationary system coherence in the 
Boltzmann equilibrated (system + environment) state for the MLSB system, 
projected onto the excited-state subspace.
Although our results are generic for the MLSB model, one context in which this ``equilibrium excited state'' problem arises is in the study of energy transfer between photosynthetic pigment-protein complexes (PPCs). In photosynthetic systems, the coupling energies \emph{between} different PPCs are typically much smaller than either the site-to-site or system-bath interaction energies \emph{within} a given PPC. As a result, thermalization {\it within} each complex occurs much more rapidly than energy transfer {\it between} complexes, and inter-PPC couplings can be treated perturbatively in a description known as multi-chromophore F{\"o}rster resonance energy transfer (MC-FRET).\cite{Ma2015} Note, however, that energy transfer dynamics \emph{within} each complex, the focus of this paper, may be far outside the F{\"o}rster regime. A natural framework for treating the time-dependence of {\it intra}complex relaxation is provided by the polaron framework which may be used to directly study the formation of stationary coherence due to the phonon-induced relaxation of the system-bath product state created by solar excitation.\cite{Olsina2014} 

Under these conditions,  the density matrix for each PPC takes the form
\begin{align}
    \hat \rho &= (1 - \pi_\text{exc} ) \hat \rho_g + \pi_\text{exc} \hat \rho_e,
\end{align}
where
\begin{align}
    \hat \rho_g &=
    \hat \rho_\text{eq}^\mathrm{B} \ket{0}\bra{0}
\\
    \label{rho_e_def}
    \hat \rho_e &=
    \frac{\hat {\mathcal P}_e \mathrm{e}^{-\beta \hat{H} } \hat {\mathcal P}_e }{\mathrm{Tr} \{ \hat {\mathcal P}_e \mathrm{e}^{-\beta \hat{H}} \hat {\mathcal P}_e \} }
\end{align}
with
\begin{align}
    \hat {\mathcal P}_e = \sum_{n=1}^{N_\mathrm{S}} \ket{n}\bra{n}
\end{align}
being the projector onto the excited state subspace, and with
\begin{align}
    \label{rho_B_eq}
    \hat \rho_\text{eq}^\mathrm{B} &=
    \frac{e^{-\beta \hat{H}_\mathrm{B}}}{\mathrm{Tr}_\mathrm{B}\{\mathrm{e}^{-\beta \hat{H}_\mathrm{B}} \}} .
\end{align}
The excited-state population $\pi_\text{exc}$ is determined by the relative rates of energy absorption and de-excitation via fluorescence or energy transfer to other complexes. 

In the absence of system-bath interactions, the equilibrium $\hat \rho_e$ would be diagonal in the eigenbasis of the system Hamiltonian
\begin{align}
    \label{qm_HS}
    \hat{H}_\mathrm{S} = \hat{H}_\mathrm{l} + \hat{H}_\mathrm{ss} ,
\end{align}
taking the form of a product between the bath equilibrium density operator $\hat \rho_\text{eq}^{\mathrm{B}}$ and the excited-state equilibrium density matrix
\begin{align}
    \label{sigma_e_0_def}
    \hat \sigma_e^{(0)} &= \hat {\mathcal P}_e \frac{ \mathrm{e}^{-\beta \hat H_e } }{Z_e^{(0)}} \hat {\mathcal P}_e ,
\end{align}
where
red{}{\begin{align}
    \label{He_quantum}
    \hat{H}_e &= \hat {\mathcal P}_e \left( \hat H_\mathrm{S} - \hbar \bar\omega \right) {\mathcal P}_e
\end{align}}
and
\begin{align}
    Z_e^{(0)} &= \text{Tr}\{ \hat {\mathcal P}_e e^{-\beta \hat H_e} \hat {\mathcal P}_e \} .
\end{align}

System-bath interactions, however, introduce correlations in  $\hat \rho_e$ both between system and bath and between different system energy states.
Correlations between system states are reflected in off-diagonal elements in the reduced system density matrix, i.e., stationary coherences
\begin{align}
    \label{Cmunu_def}
    {C}_{\mu\nu}^\mathrm{Q} &=
    \mathrm{Tr} \left \{\ket{\eigenmarker \nu} \bra{\eigenmarker \mu} \hat \rho_e \right \}, & \begin{matrix}\mu,\nu>0\\ \mu\neq\nu\end{matrix}
\end{align}
in the system eigenbasis. Here and throughout the text, Greek indices ($\mu,\nu,\kappa,\lambda,...$) are used to indicate quantities in the system eigenbasis, while Roman indices ($m,n,l,...$) indicate quantities in the local site basis. In particular, the states $\ket{\eigenmarker \mu}$ are eigenkets of $\hat{H}_e$ in the excited state manifold, with eigenvalues $\hbar \eigenmarker \omega_\mu$ and are related to the site-basis states $\ket{n}$ by
\begin{align}
    \label{u_definition}
    \ket{\eigenmarker \mu} &= \sum_m u_{\mu m} \ket{m},
\end{align}
where $u_{\mu m}$ is the real $N_\mathrm{S}\times N_\mathrm{S}$ unitary matrix that diagonalizes
$\hat{H}_e$. In this basis, $\hat{H}_e$ takes the simplified form
\begin{align}
    \hat{H}_e &= \hbar \sum_\mu \delta \eigenmarker \omega_\mu \ket{\eigenmarker \mu} \bra{\eigenmarker \mu} ,
\end{align}
where
\begin{align}
    \delta\eigenmarker \omega_\mu &= \eigenmarker \omega_\mu - \bar\omega.
\end{align}

{Two general features of equilibrium coherence in the MLSB model may be noted without detailed calculation. First, as long as the transformation matrix $\bm{u}$ is chosen to be real,
the matrix elements of the Hamiltonian are also real in the system eigenbasis, implying [see Eqs.~\eqref{rho_e_def} and \eqref{Cmunu_def}] that equilibrium coherences are also purely real. Second, equilibrium coherences vanish whenever the thermal energy $k_\mathrm{B} T = 1/\beta$ of the environment is large relative to \emph{all} the excited-state energy scales $\hbar \omega_{mn}$, $\hbar V_{mn}$ $E_{mn}^\mathrm{r}$, and $\hbar \Omega_k$. For, in this case,
\begin{align}
    \label{high_t_limit}
    \hat {\mathcal P}_e e^{-\beta \hat H} \hat {\mathcal P}_e 
    &= e^{-\beta \hbar \bar \omega} \hat {\mathcal P}_e e^{-\beta \left( \hat {\mathcal P}_e \hat H \hat {\mathcal P}_e - \hbar \bar\omega \right)} \hat {\mathcal P}_e \approx e^{-\beta \hbar \bar \omega} \hat {\mathcal P}_e ,
\end{align}
since all excited-state matrix elements of $\hat {\mathcal P}_e \hat H \hat {\mathcal P}_e - \hbar \bar\omega$ are determined by $\omega_{mn}$, $V_{mn}$, $E_{mn}^\text{r}$, and $\Omega_k$, and are thus small relative to $k_B T$. The excited-state density matrix $\hat \rho_e$ of Eq.~\eqref{rho_e_def} is therefore diagonal in the system eigenbasis, and stationary coherences [Eq. \eqref{Cmunu_def}] vanish. Physically, this result reflects the fact that the Boltzmann state assigns similar populations to states with similar energies. In the high-temperature limit (where energy differences are negligible), all excited states are thus assigned the same population. }

{However, even at 300 K, the excited-state energy scales typical of biological light-harvesting systems are comparable to $k_B T \approx 200$ cm$^{-1}$, indicating that the high-temperature limit of Eq.~\eqref{high_t_limit} is not applicable. In fact, previous studies have shown that the magnitude of the coherences ${C}_{\mu\nu}^\mathrm{Q}$ induced in $\hat \rho_e$ by system-bath interactions can be comparable to the corresponding excited-state populations.\cite{Moix2012,Lee2012,Ma2015} }
Significantly, little attention has been given to the question of whether these coherences represent a uniquely quantum-mechanical phenomenon or, alternatively, whether they can be reproduced by strictly classical models. This possibility is explored in the next section.

\section{Classical Coherence}
\label{sec_classical}
It has recently been shown\cite{Reppert2018b, Mancal2013} that light-harvesting dynamics under the MLSB model are closely mimicked by an analogous classical model with material Hamiltonian of the same form as Eq.~\eqref{def_H} with
\begin {eqnarray}
	\label{Hl_classical}
	H_\mathrm{l} &=&
	\frac {1} {2} \sum_{n= 1} ^{N_\mathrm{S}} \left (\omega_n ^ 2 q_n ^ 2 + p_n ^ 2 \right)
\\
    \label{Hss_classical}
	H_\mathrm{ss} &=&
	\sum_{m,n} \sqrt {\omega_m \omega_n } V_{mn} q_m q_n
\\
	H_\mathrm{B} &=&
	\frac {1} {2} \sum_{k} \left (\Omega_k ^ 2 Q_k ^ 2 + P_k ^ 2 \right)
\\
    \label{H_SB_C}
	H_\mathrm{SB} &=&
	\sum_{n, k} \alpha_{nk} \omega_n  q_n ^ 2 Q_k .
\end {eqnarray}
Here $q_n$ and $p_n$ are classical position and momentum coordinates for a fictitious oscillator representing the electronic motion of the $n^\text{th}$ site. As in the quantum case (Eq.~\eqref{qm_HS}), the terms $H_\mathrm{l}$ and $H_\mathrm{ss}$ together constitute the system Hamiltonian
\begin{align}
	\label{HS_cm}
	H_\mathrm{S} = H_\mathrm{l} + H_\mathrm{ss} .
\end{align}
Note that although the displacement-mediated couplings (proportional to $q_m q_n$) of Eq.~\eqref{Hss_classical} are conventional in applications to photosynthetic light harvesting,\cite{Mancal2013,Reppert2018b,Briggs2011} this form is dynamically equivalent under Eq.~\eqref{quantum_adiabatic_restriction} to the displacement/momentum coupling (proportional to $q_m q_n + p_m p_m$) more often employed in the broader context of semiclassical mappings.\cite{Meyer1979,Stock1997} (See Ref.~\cite{Briggs2012} for a detailed discussion.)

Under weak electromagnetic excitation (whether coherent or incoherent), the state of this system can be represented by an $(N_\mathrm{S}+1) \times (N_\mathrm{S}+1)$ matrix $ \densityR$, whose entries are functions of the bath coordinates $Q_k,P_k$.\cite{Reppert2018b}
The classical phase space density $\rho$ is related to $\densityR$ by
\begin{align}
    \label{classical_state_expansion}
    \rho &= \sum_{m,n=0}^{N_\mathrm{S}} R_{mn} \varrho_{mn}^\mathrm{C} ,
\end{align}
where
\begin{align}
    \label{vrhoc}
    \varrho_{mn}^\mathrm{C} &=     \sigma_\text{eq} \prod_{l=1}^{N_\mathrm{S}}
    \left(\beta \omega_l J_l\right)^{\frac{\delta_{ml} + \delta_{nl}}{2}}
    \mathrm{e}^{\mathrm{i}\left( \delta_{ml} \theta_{m} - \delta_{nl}\theta_{n}\right)} .
\end{align}
Here $\sigma_\text{eq}$ is the system Boltzmann density in the absence of the environment, and $J_n$ and $\theta_n$ are the canonical action-angle variables of the system related to $q_n$ and $p_n$ by
\begin {eqnarray}
    \label{aa_definition_q_local}
    q_n &=& \sqrt {\frac {2J_n} {\omega_n}}\cos \theta_n
\\
    \label{aa_definition_p_local}
    p_n &=& -\sqrt{2\omega_n J_n} \sin\theta_n .
\end {eqnarray}
For $m,n>0$, the functions $\varrho_{mn}^\mathrm{C}$ represent perturbative contributions to the probability density following electromagnetic  excitation of both sites $m$ and $n$. Similarly, $\varrho_{m0}^\mathrm{C} = (\varrho_{0m}^\mathrm{C})^*$ represents a perturbative contribution involving the excitation of only site $m$, while $\varrho_{00}^\mathrm{C}$ is  the equilibrium density $\sigma_\text{eq}$ with no electromagnetic field induced  excitations. The ``classical density matrix'' $\densityR$ follows a set of matrix equations that closely parallel the quantum Liouville equation for $\hat \rho$ (in the site basis) across a wide range of system-bath interactions from the  coherent Redfield regime to the incoherent F{\"o}rster regime.\cite{Reppert2018b}

As described in Ref. \cite{Reppert2018b}, site-basis coherences correspond in this classical  framework to off-diagonal elements of  the classical density matrix $\bm{R}$ after tracing out the bath. By analogy with the quantum-mechanical case, classical system eigenbasis coherences  ${C}_{\mu \nu}^\mathrm{C}$  correspond to the off-diagonal elements
\begin{align}
\label{ref32}
    {C}_{\mu \nu}^\mathrm{C} =
    \int \left( \prod_k \mathrm{d}Q_k  \mathrm{d}P_k \right ) \eigenmarker R_{\mu\nu} , && \begin{matrix}\mu,\nu > 0\\ \mu \neq \nu~, \end{matrix}
\end{align}
of the system eigenbasis ``classical density matrix''
\begin{align}
    \eigenmarker{R}_{\mu\nu} &= \sum_{m,n=1}^{N_\text{S}} u_{\mu m} u_{\nu n} R_{mn}.
\end{align}
defined for $\mu,\nu>0$.

To obtain a physical interpretation of classical coherences $C_{\mu\nu}^\mathrm{C}$, we express the expansion functions $\varrho_{mn}^\text{C}$ [Eq.~\eqref{vrhoc}] in terms of the normal mode coordinates
\begin{align}
    \eigenmarker q_\mu &= \sum_m u_{\mu m} q_m\\
    \eigenmarker p_\mu &= \sum_m u_{\mu m} p_m,
\end{align}
and obtain (for $m,n>0$)
\begin{align}
    \varrho_{mn}^\mathrm{C} 
    &= \frac{\beta}{2} \sigma_\text{eq} \left( \omega_m q_m - i p_m \right) \left( \omega_n q_n + i p_n \right) \\
    \approx& \sum_{\mu\nu} u_{\mu m} u_{\nu n} \frac{\beta}{2} \sigma_\text{eq} \left( \eigenmarker \omega_\mu \eigenmarker q_\mu - i \eigenmarker p_\mu \right) \left( \eigenmarker \omega_\nu \eigenmarker q_\nu + i \eigenmarker p_\nu \right) ,
\end{align}
and
\begin{align}
    \varrho_{m0}^\text{C} &= \left( \eigenmarker \varrho_{0m}^\text{C} \right)^* 
    = \sqrt{\frac{\beta}{2}} \sigma_\text{eq} \left( \omega_m q_m - i p_m \right) \\
    &\approx \sum_\mu u_{\mu m} \sqrt{\frac{\beta}{2}} \sigma_\text{eq} \left( \eigenmarker \omega_\mu \eigenmarker q_\mu - i \eigenmarker p_\mu \right) ,
\end{align}
where the approximations hold under Eq. \eqref{quantum_adiabatic_restriction}. A brief calculation then reveals that for $\mu \neq \nu$
\begin{align}
    \label{classical_Cre}
    \text{Re } {C}_{\mu\nu}^\mathrm{C} &=
    \frac{\left \langle \eigenmarker p_\mu \eigenmarker p_\nu \right \rangle }{k_B T} =
    \eigenmarker \omega_\mu \eigenmarker \omega_\nu \frac{\left \langle \eigenmarker q_\mu \eigenmarker q_\nu \right \rangle}{k_B T}
\\
    \label{classical_Cim}
    \text{Im } {C}_{\mu\nu}^\mathrm{C} &=
    \eigenmarker \omega_\nu \frac{\left \langle \eigenmarker p_\mu \eigenmarker q_\nu \right \rangle}{k_B T} =
    - \eigenmarker \omega_{\mu} \frac{\left \langle \eigenmarker p_\nu \eigenmarker q_\mu \right \rangle }{k_B T},
\end{align}
where the angle brackets indicate a phase-space average over the density $\rho$ in Eq.~\eqref{classical_state_expansion}.
Thus the real part of ${C}_{\mu\nu}^\mathrm{C}$ represents linear correlation between normal mode coordinates $\eigenmarker q_\mu$ and $\eigenmarker q_\nu$ or momenta $\eigenmarker p_\mu$ and $\eigenmarker p_\nu$,
while the imaginary part represents cross-correlations between position and momentum.

{It is noteworthy that an analogous result is obtained if the classical Hamiltonian [Eqs.~\eqref{Hl_classical} - \eqref{H_SB_C}] is quantized. In the single-excitation manifolds, the resulting Hamiltonian is equivalent to the quantum MLSB model of Eqs.~\eqref{Hl_quantum} - \eqref{HB_quantum} up to zero-point shifts in $\hat H_\text{SB}$ and $\hat H_\text{S}$. A brief calculation then reveals that
\begin{align}
    \label{quantum_Cre_qp}
    \text{Re } C_{\mu\nu}^\text{Q} &= \frac{\left \langle \hat p_\mu \hat p_\nu \right \rangle}{\hbar \sqrt{\omega_\mu\omega_\nu}} = \omega_\mu\omega_\nu  \frac{\left \langle \hat q_\mu \hat q_\nu \right \rangle }{\hbar \sqrt{\omega_\mu\omega_\nu}} \\
    \label{quantum_Cim_qp}
    \text{Im } C_{\mu\nu}^\text{Q} &= \omega_\nu \frac{\left \langle \hat p_\mu \hat q_\nu \right \rangle }{\hbar \sqrt{\omega_\mu \omega_\nu}} = - \omega_\mu \frac{\left \langle \hat p_\nu \hat q_\mu \right \rangle }{\sqrt{\omega_\mu \omega_\nu}}
\end{align}
where $\left \langle ... \right \rangle$ indicates a trace over the quantum density matrix $\hat \rho_e$ of Eq.~\eqref{rho_e_def} and $\hat q_\mu$ and $\hat p_\mu$ are the quantum operators associated with the classical normal mode coordinates $q_\mu$ and $p_\mu$. Thus, in both quantum and classical systems, coherences may be interpreted as indicators of correlations amongst the position and momentum coordinates of different oscillators. Note that in both cases equilibrium coherences must be purely real, so that cross-correlations between position and momentum must vanish for both quantum and classical systems. }

However, in contrast to the quantum case, which supports non-zero real components of $C_{\mu\nu}^\text{Q}$ at equilibrium, \emph{equilibrium coherences vanish entirely in the classical system, regardless of temperature and the strength of system-system and system-bath interactions}. This observation follows from the result detailed in Ref.~\cite{Reppert2018b} that the classical dynamics of $\boldsymbol{R}$ always support the coherence-free equipartition state
\begin{align}
    R_{\mu\nu} = \frac{\pi_\text{exc}}{N_\text{S}} \rho_\text{eq}^\text{B}  \delta_{\mu\nu},
\end{align}
as an equilibrium (stationary) state.\footnote{Note that the results of Ref.~\cite{Reppert2018b} do not guarantee that this equilibrium state is unique. However, thermalization to the equipartition state can be directly verified in both the opposing limits of F{\"o}rster\cite{Reppert2018b} and Redfield\cite{Mancal2013} dynamics, strongly suggesting that equipartition represents a unique equilibrium state across the complete range of system-bath interaction strengths.}
Here $\rho_\text{eq}^\text{B}$ is the equilibrium density for the isolated bath, analogous to Eq.~\eqref{rho_B_eq} for the quantum bath and the $\delta_{\mu\nu}$ ensures [see Eq.~\eqref{ref32}]  that $C_{\mu\nu}^C = 0$ for $\mu \neq \nu$.

Intuitively, this corresponds to the fact that the classical dynamics of $\boldsymbol{R}$ resemble -- at all temperatures -- those of a quantum system whose thermal energy $k_B T$ is large relative to all excited state energy scales, so that quantum commutators can be neglected.  (See Ref. \cite{Reppert2018b} and Section \ref{sec_quantumness} below.) Since equilibrium coherences vanish at high temperatures even for quantum systems (see Section \ref{sec_model}), it is perhaps unsurprising that classical coherences vanish at all temperatures.

{Physically, the failure of the classical model to capture equilibrium coherence effects may be rationalized as a consequence of the relatively small displacement in system coordinates $q_n$ during the formation of the excited states $\varrho_{mn}^\text{C}$. As discussed in Ref.~\cite{Reppert2018b}, the energy of a classical excited state $\varrho_{mn}^\text{C}$ is of the order of $k_B T$, whereas the energy of a quantum excited state $\ket{m}\bra{n}$ is of the order of $\hbar \bar\omega$, corresponding to a much larger displacement in system coordinates. (Recall that by assumption $k_B T \ll \hbar \bar\omega$.) Since the system-bath interaction strength depends (via $H_\text{SB}$) on the overall displacement of system oscillators $q_n$ away from equilibrium, it suggests that equilibrium coherence effects could be recovered by introducing energy quantization ``by hand'' into the classical model. This possibility is explored in the next section.}

\section{Semiclassical Coherence}
\label{sec_semiclassical}
To ascertain whether energy quantization alone is sufficient for non-zero equilibrium coherence formation, we employ the semiclassical framework of Refs.~\cite{Jaffe1985,Jaffe1985b,Wilkie1997a,Wilkie1997b} to map quantum states $\ket{\eigenmarker \mu}\bra{\eigenmarker \nu}$  to semiclassical states
\begin{align}
    \label{semiclassical_states}
    \eigenmarker \varrho_{\mu\nu}^\mathrm{SC} & \equiv
    \mathrm{e}^{\mathrm{i}\left( \eigenmarker \theta_{\mu} - \eigenmarker \theta_\nu \right) } \prod_{\lambda=1}^N
    \delta\left( \eigenmarker J_\lambda - \frac{\hbar}{2} \left( 1 + \delta_{\mu\lambda} + \delta_{\nu\lambda} \right) \right) .
\end{align}
Here energy quantization is enforced via the delta-function dependence of the state
$\eigenmarker \varrho_{\mu\nu}^\mathrm{SC}$ on the normal-mode action variables $\eigenmarker J_\mu$ defined via the relations (compare to the site basis expressions in  Eqs. \eqref{aa_definition_q_local} and \eqref{aa_definition_p_local})
\begin {eqnarray}
    \label{aa_definition_q_normal}
    \eigenmarker q_\mu &=& \sqrt {\frac {2\eigenmarker J_\mu} {\eigenmarker \omega_\mu}}\cos \eigenmarker \theta_\mu
\\
    \label{aa_definition_p_normal}
    \eigenmarker p_\mu &=& -\sqrt{2\eigenmarker \omega_\mu \eigenmarker J_\mu} \sin \eigenmarker \theta_\mu .
\end {eqnarray}
In the absence of the bath, diagonal states $\eigenmarker \varrho_{\mu\mu}^\mathrm{SC}$
of this form possess the same {\it classical} energies as their quantum counterparts
$\ket{\eigenmarker \mu}\bra{\eigenmarker \mu}$, while the coherence states $\eigenmarker \varrho_{\mu\nu}^\mathrm{SC}$
oscillate under the {\it classical} equations of motion with the same characteristic
frequencies $\eigenmarker \omega_{\mu \nu}$.
Moreover, in the absence of the bath, excited state matrix elements
$\braAket{\eigenmarker \mu}{\hat \rho_e}{\eigenmarker \nu}$ are given {\it exactly} for this system by the classical
phase-space integrals
\begin {align}
	\label{semiclassical_system_projections}
	\braAket{\eigenmarker \mu}{\hat \rho_e}{\eigenmarker \nu} =
	\frac{e^{\beta \left( 1 + \frac{N_\mathrm{S}}{2} \right) \hbar \bar\omega}}{Z_e^{(0)}}
	\int \left( \prod_\lambda \frac{\mathrm{d}\eigenmarker { \theta_\lambda} \mathrm{d} \eigenmarker { J_\lambda}}{2\pi} \right)
	\eigenmarker \varrho_{\nu\mu}^\mathrm{SC}\mathrm{e}^{-\beta H_\mathrm{S}},
\end {align}
where the prefactor $\mathrm{e}^{\beta \left( 1 + \frac{N_\mathrm{S}}{2} \right) \hbar \bar\omega}$ in Eq.~\eqref{semiclassical_system_projections} accounts for the $\hbar \bar\omega$ offset in $\hat{H}_e$ (compare Eq.~\eqref{He_quantum} to Eq.~\eqref{qm_HS}) and for the zero-point energy not explicitly included in $\hat{H}_\mathrm{S}$.
Thus, quantum matrix elements for an isolated system with Hamiltonian $\hat H_S$ may be calculated exactly
as projections of the classical probability density onto the appropriate semiclassical state. It is noteworthy that a similar semiclassical framework has been developed for the evaluation of nonlinear response functions, yielding exact quantum results in several important cases.\cite{Kryvohuz20015}

The significant question of interest here is whether the relationship described by Eq.~\eqref{semiclassical_system_projections} continues to hold in the presence of system-bath interactions and hence whether the quantum coherence elements ${C}_{\mu\nu}^\mathrm{Q}$ are accurately described by the semiclassical
projections
\begin {align}
	\label{semiclassical_coherence_element}
	{C}_{\mu\nu}^\mathrm{SC} =
	\frac{e^{\beta \left( 1 + \frac{N_\mathrm{S}}{2} \right) \hbar \bar\omega}}{Z_e^{(0)}}
	\int \left( \prod_\lambda \frac{\mathrm{d}\eigenmarker { \theta_\lambda} \mathrm{d} \eigenmarker { J_\lambda}}{2\pi} \right)
		\eigenmarker \varrho_{\nu\mu}^\mathrm{SC} \sigma_\text{red}^\mathrm{SC} .
\end {align}
Here the bare system density $\mathrm{e}^{-\beta H_\mathrm{S}}$ in Eq.~\eqref{semiclassical_system_projections} has been replaced by the (unnormalized) semiclassical reduced density
\begin{align}
	\label{sigma_red_sc}
	\sigma_\text{red}^\mathrm{SC} &=
	\int \left( \prod_k \mathrm{d}Q_k \mathrm{d}P_k \right)
	\frac{e^{- \beta \left( H_\mathrm{S} + H_\mathrm{SB}^\mathrm{SC} + H_\mathrm{B} \right)}}
		{Z_\mathrm{B}^\mathrm{C}},
\end{align}
where
\begin{align}
    \label{H_SB_SC}
    H_\mathrm{SB}^\mathrm{SC} &=
     \sum_{n, k} \alpha_{nk} Q_k \left( J_n - \frac{\hbar}{2} \right),
\end{align}
and where $Z_\mathrm{B}^\mathrm{C}$ is the partition function of the classical bath. 

The semiclassical Hamiltonian $H_\mathrm{SB}^\mathrm{SC}$ differs from the classical $H_\mathrm{SB}^\mathrm{C}$ in two key features. First, the quantity $\omega_n q_n^2$ in $H_\mathrm{SB}^\mathrm{C}$ has been replaced by the local-mode action $J_n = \frac{\omega_n^2 q_n^2 + p_n^2}{2\omega_n}$ in $H_\mathrm{SB}^\mathrm{SC}$, reflecting more precisely the dependence of $\hat{H}_\mathrm{SB}$ on the {\it total} energy at each local site, rather than only the {\it potential} energy associated with $q_n^2$. It is critical that this replacement does not affect the failure of the classical model to capture stationary coherence: i.e., it should be the case that a strictly classical model featuring $J_n$-coupling will fail in the same way and for the same reasons as does the classical model (with $\omega_n q_n^2$ coupling) studied in the last section. In fact, under the restrictions in Eqs. \eqref{quantum_adiabatic_restriction} and \eqref{quantum_temperature_restriction}, the classical dynamics are unaffected by the form of the coupling (see Eq.~A15 of Ref.\cite{Reppert2018b}). The replacement does, however, affect semiclassical matrix elements; due to its closer connection to the quantum Hamiltonian, the semiclassical Hamiltonian of Eq. \eqref{H_SB_SC} gives semiclassical matrix elements in better agreement with their quantum analogs.

Second, a zero-point offset of $\frac{\hbar}{2}$ has been added in $H_\mathrm{SB}^\mathrm{SC}$ to ensure that (as in the quantum model) system-bath interactions vanish in the semiclassical ground state. Note again that since this replacement is equivalent to an overall shift of the bath coordinates (and thus can be eliminated by simply redefining the zero-point of the bath modes), it has no impact on the results of the last section.  Thus, despite the differences between Eqs. \eqref{H_SB_C} and \eqref{H_SB_SC}, a strictly classical treatment of either model will (as in Section \ref{sec_classical}) fail to capture equilibrium coherence. The next section examines whether this limitation is remedied by enforcing energy quantization via the semiclassical matrix elements of Eq. \eqref{semiclassical_coherence_element}.
%

\subsection{Coupled Dimer}
For simplicity, consider now the case $N_\mathrm{S}= 2$ of a coupled dimer.
By completing the square in the exponent, the bath integration in Eq. \eqref{sigma_red_sc} can be carried out explicitly to obtain
\begin{align}
\sigma_\text{red}^\mathrm{SC} &\approx\mathrm{e}^{-\beta ( H_\mathrm{S} + H_\text{eff} ) }
\end{align}
with the effective Hamiltonian $H_\text{eff}$
\begin{align}
    H_\text{eff} 
    &= - \sum_{mn} E_{mn}^\mathrm{r} \left( \frac{J_m}{\hbar} - \frac{1}{2} \right) \left( \frac{J_n}{\hbar} - \frac{1}{2} \right)
\end{align}
Significantly, the semiclassical reduced density $\sigma_\text{red}^\text{SC}$ is seen to be determined completely by the reorganization energies $E_{mn}^\mathrm{r}$ . Here $H_\text{eff}$ is a function of the local actions $J_n = \frac{1}{2 \omega_n }(\omega_n^2 q_n^2 + p_n^2)$ that are related to the
normal mode coordinates by
\begin{align}
    \label{Jn_exact}
    J_n &= \frac{1}{2} \sum_{\mu\nu} u_{\mu n} u_{\nu n}
    \left( \omega_n \eigenmarker q_\mu \eigenmarker q_\nu + \frac{\eigenmarker p_\mu \eigenmarker p_\nu}{\omega_n} \right)
\\
    \label{Jn_simple}
    &\approx \sum_{\mu\nu} u_{\mu n} u_{\nu n} \sqrt{\eigenmarker J_\mu \eigenmarker J_\nu} \cos \left( \eigenmarker \theta_{\mu} - \eigenmarker \theta_{\nu} \right) .
\end{align}
Eq.~\eqref{Jn_simple} is obtained from Eq.~\eqref{Jn_exact} by expanding $\eigenmarker q_\mu$ and $\eigenmarker p_\mu$ in action-angle variables and noting that $\frac{\omega_n}{\sqrt{\eigenmarker \omega_\mu \eigenmarker \omega_\nu}} \approx 1$ under Eq.~\eqref{quantum_adiabatic_restriction}.

For the coupled dimer, we have simply
\begin{align}
    \bm{u} &= \left [ \begin{matrix} \cos \phi & -\sin \phi \\ \sin \phi & \cos \phi \end{matrix} \right]
\end{align}
with
\begin{align}
    \tan \phi = \frac{2 V_{12}}{\Delta + \sqrt{\Delta^2 + 4 V_{12}^2 }} ,
\end{align}
in terms of the site-basis frequency difference
\begin{align}
    \Delta = \omega_{n=2} - \omega_{n=1} .
\end{align}

Defining the angle difference
\begin{align}
    \Theta = \theta_{\mu=2} - \theta_{\mu=1},
\end{align}  Eq.~\eqref{semiclassical_coherence_element} now reads
\begin{align}
    \label{Cmunu_SC}
    {C}_{12}^\mathrm{SC} &= \frac{1}{Z_e^{(0)}} \int_0^{2\pi} \left( \prod_\lambda \frac{\mathrm{d}\eigenmarker \theta_\lambda}{2\pi} \right)
    \mathrm{e}^{-\beta H_\mathrm{eff}^\mathrm{o}}\mathrm{e}^{\mathrm{i} \Theta}
\end {align}
where
\begin{align}
    & H_\mathrm{eff}^\mathrm{o} = - 2 E_{12}^\mathrm{r}
    \left( \frac{1}{4} - 4 f_\phi^2 \cos^2 \Theta \right)
\\
    & - E_{11}^\mathrm{r} \left( \frac{1}{2} + 2 f_\phi \cos \Theta \right)^2 -
    E_{22}^\mathrm{r} \left( \frac{1}{2} - 2 f_\phi \cos \Theta \right)^2 \nonumber
\end{align}
with
\begin{align}
    f_\phi = \cos \phi \sin \phi.
\end{align}
$H_\mathrm{eff}^\mathrm{o}$ is simply $H_\text{eff}$ evaluated at $\eigenmarker J_\mu = \hbar$ for all $\mu$.

Eq.~\eqref{Cmunu_SC} is difficult to evaluate in general. However, a few general observations are possible. Noting that $H_\mathrm{eff}^\mathrm{o}$ is a function only of $\cos \Theta$ implies that the integral is purely real, i.e. that
\begin{align}
    \label{semiclassical_coherence_dimer_exact}
    {C}_{12}^\mathrm{SC} &= \frac{1}{Z_e^{(0)}} \int_0^{2\pi} \left( \prod_\lambda \frac{\mathrm{d}\eigenmarker \theta_\lambda}{2\pi} \right)
    \mathrm{e}^{-\beta H_\mathrm{eff}^\mathrm{o}}\cos \Theta .
\end {align}
Moreover, if $E_{11}^\mathrm{r} = E_{22}^\mathrm{r}$ so that $H_\mathrm{eff}^\mathrm{o}$ depends only on $\cos^2 \Theta$, the integrand is overall an odd-order polynomial in $\cos \Theta$, and again ${C}_{12}^\mathrm{SC} = 0$.

Beyond these preliminary observations, explicit evaluation of
Eq.~\eqref{semiclassical_coherence_dimer_exact} is difficult in general.
An approximate expression is readily obtained, however, by expanding the exponential
to first order. The resulting expression is accurate to second order in $H_\mathrm{SB}$
and reads
%
\begin{align}
    \label{semiclassical_coherence_dimer_approx}
    {C}_{12}^\mathrm{SC} &=
    \frac{\beta}{Z_e^{(0)}} \cos \phi \sin \phi \left( E_{11}^\mathrm{r} - E_{22}^\mathrm{r} \right)
\end {align}
\emph{Thus, to second order, semiclassical coherences are directly proportional to the
difference in reorganization energies associated with the two sites.}
Semiclassical coherences are largest at low temperatures and for strong delocalization. In the high-temperature limit $\beta \to 0$ (or alternatively in the classical limit $\hbar \to 0$ and, hence, $E_{mn}^\mathrm{r} \to 0$ [see Eq.~\eqref{lambda_definition}]), the semiclassical result approaches the classical limit $C_{12}^\mathrm{C} = 0$ obtained in the last section.

\section{Quantum Coherence}
\label{sec_quantum}
To enable a quantitative comparison between semiclassical and quantum predictions,
a similar second-order expansion can be carried out for the quantum reduced density
matrix $\hat \sigma_e = \mathrm{Tr}_\mathrm{B}\{\hat \rho_e \} $.
Expanding the quantum exponential $\mathrm{e}^{-\beta \hat{H}}$ of Eq.~\eqref{rho_e_def} to second order
in $\hat{H}_\mathrm{SB}$ gives\cite{Geva2000,Lee2012}
\begin{align}
\label{sigma_e_second_order}
\hat \sigma_e &\approx \hat \sigma_e^{(0)} \left( 1 - Z_e^{(2)} \right) + \hat \sigma_e^{(2)}
\end{align}
where
\begin{align}
Z_e^{(2)} &= \mathrm{Tr}_\mathrm{S} \left \{ \hat \sigma_e^{(2)} \right \},
\end{align}
and
\begin{align}
    \label{sigma_e_2_def}
	\hat \sigma_e^{(2)} &=
	\hbar^3 \hat \sigma_e^{(0)} \int_0^\beta \mathrm{d} s_2 \int_0^{s_2} \mathrm{d} s_1
	\sum_{mn=1}^{N_\mathrm{S}} \sum_{k} \frac{\alpha_{mk} \alpha_{nk}}{2\Omega_k}
\\
	&\times \left(\mathrm{e}^{(s_2 - s_1) \hbar \Omega_k } {\bar n}(\Omega_k) -
	\mathrm{e}^{ - (s_2 - s_1) \hbar \Omega_k } {\bar n}(- \Omega_k) \right)
\nonumber  \\
	& \quad \times \mathrm{e}^{s_2 \hat{H}_e} \ket{m}\bra{m}\mathrm{e}^{( s_1 - s_2)
	\hat{H}_e} \ket{n}\bra{n}\mathrm{e}^{- s_1 \hat{H}_e}
\nonumber
\end{align}
with
\begin{align}
{\bar n}(\Omega) &= \frac{1}{e^{\beta \hbar \Omega} - 1 } .
\end{align}
(Recall that $\hat\sigma_e^{(0)}$ was defined in Eq.~\eqref{sigma_e_0_def}.) Eqs.~\eqref{sigma_e_second_order} can be derived by setting $\hat X = \hat H_e + \hat{\mathcal P}_e \hat H_\mathrm{B} \hat{\mathcal P}_e $ and $\hat Y = \hat H_\mathrm{SB} $ in the operator expansion
\begin{align}
    \label{exponential_identity}
    e^{-\beta\left( \hat X + \hat Y \right) } =& e^{-\beta \hat X} \sum_{n=0}^\infty  \int_0^{\beta} ds_n \int_0^{s_n} ds_{n-1} ... \int_0^{s_2} ds_1 \nonumber \\
    \times (-1)^n &\left( e^{  s_n \hat X} \hat Y e^{- s_n \hat X} \right) ... \left( e^{ s_1 \hat X} \hat Y e^{ - s_1 \hat X} \right),
\end{align}
truncating at second order, and tracing over the bath. The expansion [Eq. \eqref{exponential_identity}] may be verified directly by differentiating both sides to show that they satisfy the same operator differential equation and boundary conditions.

Expanding in the eigenbasis states via the relation
\begin{align}
    \ket{m} = \sum_\mu u_{\mu m} \ket{\eigenmarker \mu} ,
\end{align}
the integral in Eq.~\eqref{sigma_e_2_def} may
be evaluated directly to obtain the second-order density matrix
\begin{align}
\label{equ:rho_quantum_second_order}
	\hat \sigma_e^{(2)}&= \frac{\mathrm{e}^{-\beta \hbar \frac{\delta \eigenmarker \omega_{\mu} + \delta \eigenmarker \omega_\nu}{2}}}{Z_e^{(0)}}
	\sum_{\mu \nu \kappa} \sum_{mn}
	u_{\mu m} u_{\kappa m} u_{ \kappa n} u_{\nu n}
\nonumber \\
	& \times \int \mathrm{d}\Omega {\mathcal J}_{mn} (\Omega) \bar n(\Omega) K_\kappa^{\mu\nu}(\Omega)
	\proj{\eigenmarker \mu}{\eigenmarker \nu}
\end{align}
where we have defined an antisymmetric spectral density
\begin{align}
    {\mathcal J}_{mn}(\Omega) &= \hbar \sum_k \frac{\alpha_{nk} \alpha_{mk}}{2\Omega_k}
    \left [ \delta(\Omega - \Omega_k) - \delta(\Omega + \Omega_k) \right ]
\end{align}
and the integration kernel
\begin{align}
    \label{K_kappa_def}
    K_\kappa^{\mu\nu} &=
    \frac{e^{- \frac{\beta \hbar \eigenmarker \omega_{\mu\nu}}{2} }}{\eigenmarker \omega_{\mu\nu}
    \left( \Omega + \eigenmarker \omega_{\mu\kappa} \right) } -
    \frac{e^{ \frac{\beta \hbar \eigenmarker \omega_{\mu\nu}}{2}}}{\eigenmarker \omega_{\mu\nu}
    \left( \Omega + \eigenmarker \omega_{\nu\kappa} \right) }
\\
    &\quad\quad+ \frac{\mathrm{e}^{ \beta \hbar
    \left( \Omega - \eigenmarker \omega_\kappa + \frac{\eigenmarker \omega_\mu +
    \eigenmarker \omega_\nu}{2} \right) }}{\left( \Omega + \eigenmarker \omega_{\nu\kappa} \right)
    \left( \Omega + \eigenmarker \omega_{\mu\kappa} \right) } .
\nonumber
\end{align}
For diagonal elements ($\mu = \nu$), Eq.~\eqref{K_kappa_def} must be understood as the limiting value as $\omega_\mu \to \omega_\nu$.

The coherences are obtained from Eq.~\eqref{equ:rho_quantum_second_order} as
\begin{align}
\label{cMuNu_quantum_second_order}
	{C}_{\mu\nu}^\mathrm{Q} &= \left( \hat \sigma_e^{(2)} \right)_{\mu \nu} .
\end{align}
Expanding the kernel $K_\kappa^{\mu\nu}$ and the occupation number $\bar n(\Omega)$ in small $\beta$ reveals that in the high-temperature limit $\beta \to 0$, the semiclassical
[Eq.~\eqref{semiclassical_coherence_dimer_approx}] and quantum
[Eq.~\eqref{cMuNu_quantum_second_order}] second-order expressions become identical,
approaching (in the case of a dimer) the common limit\footnote{Note in the semiclassical case that $Z_e^{(0)} \to 2$ as $\beta \to 0$, representing equal population of both excited states.}
\begin{align}
    \label{Cmunu_second_Order_limit}
    {C}_{12}^\mathrm{SC} &\approx \frac{\beta}{2} \cos \phi \sin \phi \left( E_{11}^\mathrm{r} - E_{22}^\mathrm{r} \right) .
\end {align}

\begin{figure}
    \centering
    \includegraphics{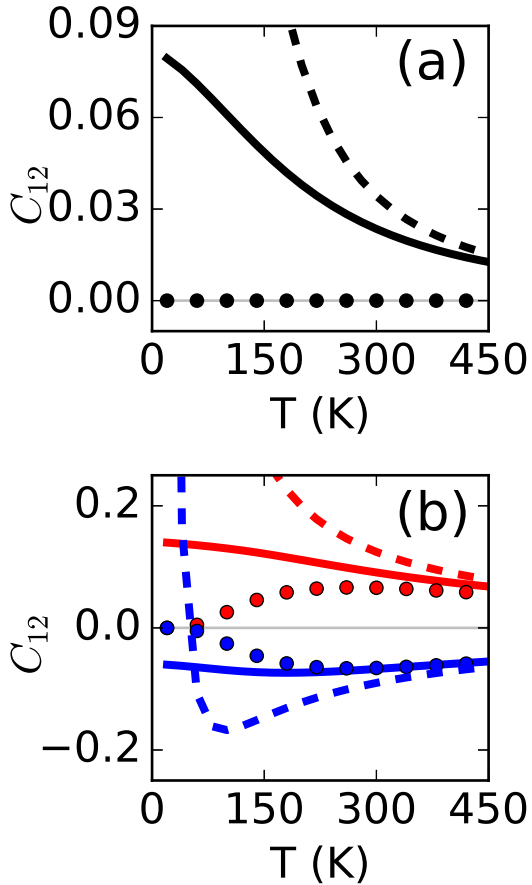}
    \caption{Comparison of classical (thin grey lines), semiclassical (dots), and quantum (thick lines) equilibrium coherences $C_{12}$ for a coupled dimer with (frame (a)) identical,
    uncorrelated system-bath interactions at each site or with (frame (b)) system-bath
    interactions at only site 1 (lower frequency, red curves) or only site 2 (higher frequency, blue curves).
    In all cases, $V_{12} = \Delta = 2\pi c \cdot 200 \text{ cm}^{-1}$, while the
    reorganization energy $E_{nn}^\mathrm{r}$ was set to $2\pi c \hbar \cdot 100 \text{ cm}^{-1}$
    for each site that interacts with the environment.
    Dashed curves indicate approximate quantum matrix elements from the third-order expansion of Eq.~\eqref{C12_third_order}.}
    \label{fig_comparison}
\end{figure}

For finite temperatures, however, quantum and semiclassical coherences have quite
distinct properties.
In particular, in contrast to the semiclassical case (where only the diagonal terms $E_{nn}^\mathrm{r}$ contribute [Eq.~\eqref{semiclassical_coherence_dimer_approx}] to equilibrium coherence) \emph{the degree of correlation between
system sites -- i.e., the form of ${\mathcal J}_{mn}(\Omega)$ when $m\neq n$ -- is crucial in the quantum system.}
For concreteness, suppose ${\mathcal J}_{mn}(\Omega) = c{\mathcal J}(\Omega)$ for all $m\neq n$, with
$-1 \leq c \leq 1$.\footnote{For a dimer, this form includes all possible cases where system-bath
interactions are symmetric, i.e., where ${\mathcal J}_{11}(\Omega) = {\mathcal J}_{22}(\Omega)$.} The case $c = 1$ corresponds to perfect correlation between the sites ($\alpha_{nk} = \alpha_{mk}$), while $c = -1$ represents perfect anticorrelation (i.e., $\alpha_{1k} = -\alpha_{2k}$) and $c = 0$ a complete lack of correlation (i.e., independent baths).
Eq.~\eqref{cMuNu_quantum_second_order} then takes the form
\begin{align}
{C}_{\mu\nu}^\mathrm{Q} &=
	(1 - c) \frac{e^{-\beta \hbar \frac{\delta \eigenmarker \omega_{\mu}
	+ \delta \eigenmarker \omega_\nu}{2}}}{Z_e^{(0)}} \sum_{n\kappa} u_{\mu n} u_{\nu n} u_{\kappa n}^2
\nonumber \\
	& \times \int \mathrm{d}\Omega {\mathcal J} (\Omega) \bar n(\Omega) K_\kappa^{\mu\nu}(\Omega).
\end{align}
For perfect correlation ($c = 1$, where $E_{mn}^\mathrm{r}$ is independent of $m$ and $n$), stationary coherences vanish, just as in the semiclassical system.
In the absence of perfect correlation, however, stationary coherence persists in the quantum
system and is maximized for anticorrelated interactions ($c = -1$). In contrast, the corresponding semiclassical coherences vanish identically, regardless of the value of $c$.

These findings emphasize the essential distinction between correlated and anticorrelated system-bath interactions in the MLSB model.\cite{Tiwari2013} Since all system energy levels respond identically to the fluctuations of a correlated bath ($c = 1$, $\alpha_{mk} = \alpha_{nk}$), energy \emph{differences} between system eigenstates are unaffected by bath dynamics and (although the system eigenvalues may fluctuate) the system eigen\emph{states} are static in time. Correlated system-bath interactions are thus incapable of driving electronic energy transfer\cite{Tiwari2013} or, as demonstrated here, of inducing stationary coherence in the equilibrium MLSB model. In contrast, \emph{anticorrelated} system-bath interactions cause strong variations in intra-system energy differences since bath dynamics induce anticorrelated fluctuations in local site energies. Thus anticorrelated vibrations strongly distort the energetic structure of the system, triggering both nonadiabatic electronic energy transfer\cite{Tiwari2013} and, as seen here, the formation of stationary coherence.

For illustration, Figure~\ref{fig_comparison} plots second-order quantum (thick solid lines) and semiclassical (dots) calculations for $C_{12}$ for a coupled dimer with $V_{12} = \Delta = 2\pi c \cdot 200 \text{ cm}^{-1}$. The spectral densities are taken to have an Ohmic form with exponential cutoff $\mathcal{J}_{mn}(\Omega) \propto \Omega \mathrm{e}^{-
\Omega / \Omega_\mathrm{c}}$ with $\Omega_c = 2 \pi c \cdot 50$ cm$^{-1}$. The classical result $C_{12}^\text{C} = 0$ is indicated by a thin grey line.
In Frame (a), system-bath interactions are taken to be identical but uncorrelated between the two sites, with $E_{mn}^\mathrm{r} = \delta_{mn} 2\pi c \hbar \cdot 100 \text{ cm}^{-1}$.
In Frame (b), interactions are localized at either site 1 (lower frequency, red curves, $E_{mn}^\mathrm{r} = \delta_{m1} \delta_{n1} 2\pi c \hbar \cdot 100 \text{ cm}^{-1}$) or site 2 (higher frequency, blue curves, $E_{mn}^\mathrm{r} = \delta_{m2} \delta_{n2} 2\pi c \hbar \cdot 100 \text{ cm}^{-1}$). Note that in both cases the magnitude of the quantum equilibrium coherence can be substantial relative to the excited-state populations (which sum to unity). In Frame (b), where only a single system site couples to the bath, the semiclassical results provide a reasonable approximation to the full quantum expression at high temperatures (above 300 K), although, in contrast to the quantum case, semiclassical coherences decay to zero at low temperatures. In contrast, the semiclassical result fails at all temperatures to capture the quantum coherence exhibited in Frame (a), corresponding to configurations in which both system sites couple with equal strength to the thermal environment. This last failure is particularly noteworthy in that system-bath interaction strengths are typically assumed to be similar at all sites in biological light-harvesting systems, suggesting that semiclassical descriptions are unlikely to capture equilibrium coherence effects in these systems.

\section{Quantum Perturbation Expansion}
\label{sec_hbar_expansion}
The fact that even semiclassical descriptions can, as seen in Figure \ref{fig_comparison}, capture some features of quantum stationary coherence at physiological temperatures suggests that a perturbative expansion in ``quantumness'' could prove computationally useful under biologically-relevant conditions. A convenient route to such an expansion is to use the Zassenhaus formula\cite{Casas2012} to expand the exponential $e^{-\beta \left( \hat H_e + \hat H_\mathrm{SB} + \hat H_\mathrm{B} \right)}$ and then to perform a partial Wigner transformation\cite{Hillery1984} over the bath coordinates. As detailed in Appendix \ref{quantum_perturbation_expansion_appendix}, this procedure yields, to order $\hbar^3$,
\begin{widetext}
\begin{align}
    \label{C12_hbar3}
    {C}_{\mu\nu}^\mathrm{Q} &=
    \frac{1}{N_\mathrm{S}}  \left \langle \braAket{\mu}{ \frac{\beta^2}{2} \left(\hat{H}_\mathrm{SB}^\mathrm{QC}\right)^2 -
    \frac{\beta^3}{6} \left( \hat{H}_\mathrm{SB}^\mathrm{QC} \hat{H}_e \hat{H}_\mathrm{SB}^\mathrm{QC} +
    \left( \hat{H}_\mathrm{SB}^\mathrm{QC}\right)^2 \hat{H}_e +
    \hat{H}_e \left( \hat{H}_\mathrm{SB}^\mathrm{QC}\right)^2 \right) }{\nu} \right\rangle_\mathrm{B} + (\hbar^4),
\end{align}
\end{widetext}
where $\hat H_\mathrm{SB}^\mathrm{QC}$ is the system operator obtained by replacing the bath operator $\hat Q_k$ with the classical coordinate $Q_k$ in Eq.~\eqref{HSB_quantum}, and where the notation $\left \langle ... \right \rangle_B$ indicates a \emph{classical} average over the equilibrium bath ensemble. For the special case of a dimer where
\begin{align}
    \delta \eigenmarker \omega_{\mu=1} &= - \delta \eigenmarker \omega_{\mu=2},
\end{align}
the two $\left( \hat H_\mathrm{SB}^\mathrm{QC} \right)^2$ terms in Eq.~\eqref{C12_hbar3} cancel leaving, after a brief calculation
\begin{align}
    \label{C12_third_order}
    {C}_{12}^\mathrm{Q} 
    &= \frac{\beta}{2} \cos \phi \sin \phi \left( E_{11}^\mathrm{r} - E_{22}^\mathrm{r} \right)
\nonumber \\
    &+ \frac{\beta^2}{12} \hbar \Delta_\mathrm{S}  \cos \phi \sin \phi (\cos^2 \phi - \sin^2 \phi)
\nonumber \\
    &\quad \quad \times\left( E_{11}^\mathrm{r} + E_{22}^\mathrm{r} - 2 E_{12}^\mathrm{r} \right)  + \mathcal{O}(\hbar^4),
\end{align}
where
\begin{align}
    \Delta_\mathrm{S} &= \omega_{\mu=2} - \omega_{\mu=1}
\end{align}
is the frequency difference between the two system eigenstates. The first term here (of order $\hbar^2$ since $E_{mn}^\mathrm{r} = \hbar^2 \sum_k \frac {\alpha_{mk} \alpha_{nk} } {2 \Omega_k^2 }$) is exactly the mutual high-temperature limit [Eq.~\eqref{Cmunu_second_Order_limit}] of the semiclassical and quantum second-order expansions. As explored already, this $\hbar^2$ term vanishes for symmetric coupling strengths ($E_{11}^\mathrm{r} = E_{22}^\mathrm{r}$). This failing is corrected by the $\hbar^3$ term in the last line of Eq. \eqref{C12_third_order} which breaks the symmetry between the two local sites and produces non-zero stationary coherence elements even for symmetric coupling strengths.

For illustration, stationary coherence elements $C_{12}^\mathrm{Q}$ calculated using
Eq.~\eqref{C12_third_order} are plotted as a function of temperature in
Figure~\ref{fig_comparison} (dashed lines) for the dimer parameters considered above.
At high temperatures, the third-order result captures the correct qualitative behavior for both symmetric (Frame a) and asymmetric (Frame b) system-bath interactions.
At low temperatures, however, the expansion fails due to the finite power of $\beta = \frac{1}{k_\mathrm{B} T}$. The cross-over regime in which the expansion begins to reasonably approximate the true quantum result occurs near ambient temperatures ($T = 300$ K), reflecting the fact that (for our model parameters) it is in this region that the thermal energy ($k_B T \approx 200 $ cm$^{-1}$ at 300 K) becomes comparable to the energy scales $\hbar \Delta$, $E_{nn}^\mathrm{r}$, $\hbar V_{12}$, and $\hbar \Omega_k$. More generally, low-order $\hbar$ expansions are expected to become accurate whenever $k_B T$ becomes comparable to or greater than all excited-state energy scales ($\hbar\omega_{mn}$, $V_{mn}$, $E_{mn}^\text{r}$, and $\hbar \Omega_k$).

These findings indicate that, although quantum finite  $\hbar$ expansions will perform poorly at low temperatures, they may be very useful for applications at ambient temperatures. This approach may be particularly relevant to biological systems, where reorganization energies are typically assumed to be comparable at all sites so that the semiclassical treatment of Section \ref{sec_semiclassical} fails completely to capture stationary coherence effects. A further advantage of this $\hbar^n$ expansion is that quantum properties are expressed [as in Eq. \eqref{C12_hbar3}] in terms of \emph{strictly classical} bath averages and thus may be amenable to evaluation using standard classical treatments such as molecular dynamics.

\section{A Phase-Space Comparison}
\label{sec_phase_space}

\begin{figure*}[htb]
    \centering
    \includegraphics{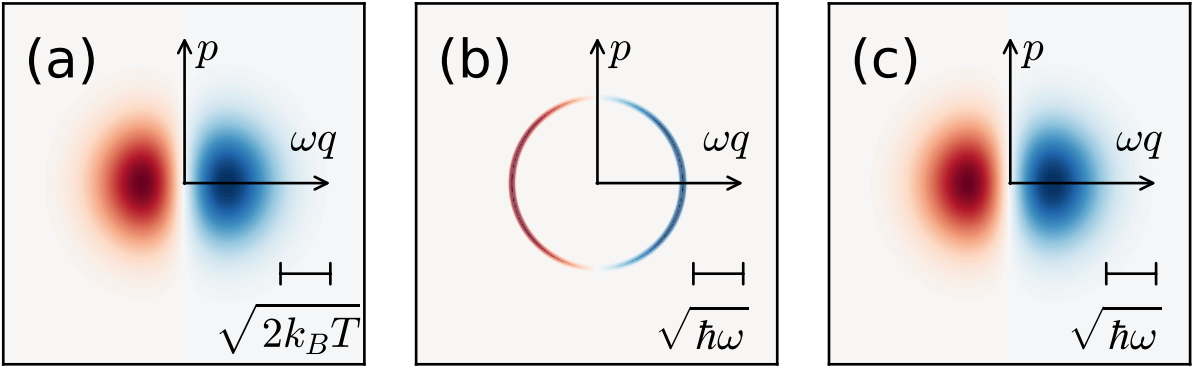}
    \caption{The real part of the classical [frame (a)], semiclassical [frame (b)], and quantum [frame (c)] phase-space distributions $\varrho_{10}$ for a single harmonic oscillator defined by Eqs.~\eqref{rho10_c} - \eqref{rho10_q}. Positive regions are shaded blue, while negative regions are shaded red. Each plot is normalized to have a maximum amplitude of unity. For visualization purposes, the denta function in $\varrho_{10}^\mathrm{SC}$ is assigned a finite width of $0.1\sqrt{\hbar\omega}$.}
    \label{fig_densities}
\end{figure*}

The results above provide insight into the origin of equilibrium stationary coherence in classical, semiclassical, and quantum descriptions of the MLSB  model.
To understand these results physically, it is useful to cast all three models into a phase-space representation where the properties of coherences can be compared directly.
Classical phase-space coherences are  defined by Eq.~\eqref{vrhoc}, while (as discussed in detail in Ref.\cite{Jaffe1985}) semiclassical coherences are defined in phase space by Eq.~\eqref{semiclassical_states}.
A phase-space representation of the quantum coherence elements $\ket{\eigenmarker \mu}\bra{\eigenmarker \nu}$ can be obtained using a Wigner transformation over the system coordinates.
Quantization of the classical system Hamiltonian $H_\mathrm{S}$ of Eq.~\eqref{HS_cm}
produces a quantum harmonic oscillator Hamiltonian that (apart from a zero-point
offset) is identical in the ground-plus-single-excitation subspace to the quantum
operator $\hat{H}_\mathrm{S}$ of Eq.~\eqref{qm_HS}.
A phase-space representation
$\varrho_{\mu\nu}^\mathrm{Q} = \left [ \ket{\eigenmarker \mu} \bra{\nu} \right]_W$
of the coherence $\ket{\eigenmarker \mu}\bra{\eigenmarker \nu}$ of this harmonic model is
then readily obtained by applying the appropriate raising and lowering operators
to the phase-space representation of the quantum ground state $\ket{0}\bra{0}$.\cite{Hillery1984} (See Appendix \ref{sec_rho_10_W}.)

As a simple example, Figure~\ref{fig_densities} plots classical [frame (a)],
semiclassical [frame (b)], and quantum [frame (c)] phase-space distributions
$\varrho_{10}$ for a single harmonic oscillator.
The distributions are defined in this case by
\begin{align}
    \label{rho10_c}
    \varrho_{10}^\mathrm{C} &=
    \frac{\omega}{2 \pi k_\mathrm{B} T} \left( \frac{ \omega q - \mathrm{i} p }{\sqrt{2 k_\mathrm{B} T}} \right)
    \mathrm{e}^{-\frac{\omega^2 q^2 + p^2}{2 k_\mathrm{B} T}}
\\
    \label{rho10_sc}
    \varrho_{10}^\mathrm{SC} & = \left( \frac{ \omega q - \mathrm{i} p }{\sqrt{2\hbar \omega}} \right) \delta\left( \frac{\omega^2 q^2 + p^2 }{2 \omega} - \hbar \right)
\\\
    \label{rho10_q}
    \varrho_{10}^\mathrm{Q} &=
    \frac{2}{\pi\hbar} \left( \frac{ \omega q - \mathrm{i} p }{\sqrt{2\hbar \omega}} \right) \mathrm{e}^{- \frac{ \omega^2 q^2 + p^2}{\hbar \omega} } .
\end{align}
In each case, only the real part of the distribution is plotted; the imaginary
part is similar but rotated by 90$^\text{o}$.
Blue represents positive features, while red represents negative; each plot
is normalized to have a maximum amplitude of unity.
For visualization purposes, the delta function in $\varrho_{10}^\mathrm{SC}$
is  assigned a finite width of $0.1 \sqrt{\hbar\omega}$.

It is noteworthy that, in all three descriptions, the variation of $\varrho_{10}$
with the phase-space angle $\theta$ is determined by the factor
\begin{align}
    \omega q - \mathrm{i} p \propto\mathrm{e}^{\mathrm{i}\theta} .
\end{align}
Thus the $\varrho_{10}$ coherence state is  associated with the displacement of phase-space amplitude along either the $q$ ($\text{Re} \varrho_{10}$) or $p$ ($\text{Im} \varrho_{10}$) axes.
Since, in the multi-oscillator case, the phase-space representation of $\ket{\eigenmarker \mu}\bra{\eigenmarker \nu}$ is found by the same procedure to be proportional to the product $\left( \omega_\mu q_\mu - \mathrm{i} p_\mu \right) \left( \omega_\nu q_\nu + \mathrm{i} p_\nu \right)$, this confirms the suggestion of Eqs.~\eqref{classical_Cre} - \eqref{quantum_Cim_qp} that coherences are associated in phase space with the correlated displacement of two distinct modes.

Beyond this shared rotational symmetry, however, the various phase-space
representations differ considerably.
That is, although the quantum and classical distributions have
identical forms the width of the distribution is
proportional to $\sqrt{2 k_\mathrm{B} T}$ in the classical system and
$\sqrt{\hbar \omega}$ in the quantum system.
As noted earlier, this difference in scale is
 the essential reason why the classical model fails
to capture stationary coherence effects in excited-state equilibration
for the MLSB model.
Whereas photoexcitation creates a comparatively large displacement in
the quantum oscillator, it induces a much smaller displacement for the
classical oscillator.
Since the system-bath interaction Hamiltonian scales with $q^2$, this
implies that the excited quantum system experiences a much stronger
interaction with the environment than does the excited classical system.
Hence, the quantum system exhibits stationary
coherence, while the classical system does not.

On the other hand, the semiclassical distribution plotted in Frame \ref{fig_densities}(b)
has the same overall dimensions in phase-space as the quantum distribution (see scale bar at lower right of each panel).
Formation of such a semiclassical excited state thus induces system-bath
interactions of the same magnitude as the quantum system, giving rise, as
seen earlier, to nonvanishing stationary coherence elements.
Nonetheless, the actual shape of the semiclassical distribution differs strongly
from that of the quantum density, so  that the two models differ
in their detailed predictions for stationary coherence elements.
It is encouraging, however, that the simple extension of this semiclassical theory
represented by Eq.~\eqref{C12_third_order} offers, as seen in Figure \ref{fig_comparison}, a fairly accurate description
of equilibrium stationary coherence at biologically-relevant temperatures, at least in the
weak-coupling limit for which the quantum expression
[Eq.~\eqref{cMuNu_quantum_second_order}] is valid.

\section{Is Equilibrium Coherence Quantum?}
\label{sec_quantumness}

Finally, let us return to the original question:  is equilibrium coherence in the MLSB
model an intrinsically quantum-mechanical effect?
The results of Section~\ref{sec_classical} indicate that equilibrium coherence cannot
be described by existing classical models.
Section~\ref{sec_semiclassical} show that semiclassical principles are capable of
capturing equilibrium coherence in certain circumstances, namely at high temperatures
and in the presence of strong asymmetry between system-bath coupling strengths
at different system sites.
However, even this semiclassical treatment fails qualitatively in other cases, i.e., for configurations in which system-bath interactions have the same magnitude at all system sites.
This last failing is particularly noteworthy in that it is not an obvious temperature effect and that system-bath interactions are typically expected to be similar for all sites in a photosynthetic PPC.
Thus, although we cannot exclude the possibility that better classical descriptions could be developed, our results suggest that (at least in photosynthetic systems)  \emph{equilibrium coherence represents a uniquely quantum-mechanical phenomenon. }

From an applications perspective, these findings are significant in that they suggest that equilibrium coherence effects may be an exclusive feature of quantum devices.
This is in stark contrast to dynamical coherence effects which can be exploited in both classical and quantum architectures.\cite{Leon2013,Leon2015}
Thus, in the search for novel quantum devices, stationary coherence effects appear to
be a promising area of research.
It should be emphasized, however, that our present findings apply only to equilibrium effects in the MLSB model.
It remains to be seen whether the same results extend to more general nonequilibrium steady-state processes and/or to other model systems. %

It is also worth emphasizing that the failure of the semiclassical framework of Section \ref{sec_semiclassical} to capture stationary coherence effects does not exclude the possibility that other semiclassical methods might be more successful. Although our calculations demonstrate that quantization of the system action is not alone sufficient, it would be of interest to explore whether other semiclassical approximations (see, for some examples, Refs.\cite{Miller2001,Habershon2013,Kapral2015}) perform better and, if so, what physical features are essential to the generation of equilibrium coherence. In comparison to other methods, it is noteworthy that the semiclassical framework of Section \ref{sec_semiclassical} is based on a static identification of quantized-action system states and does not specify a concrete time-evolution propagator for the system+bath composite.\footnote{Note that the approach of Jaff{\'e} and Brumer specifies equations of motion only for closed systems.\cite{Jaffe1985,Jaffe1985b} In principle, the approach could yield dynamical equations for the entire system+bath composite, but only if classical action/angle variables could be identified for the entire supersystem, a task which is greatly complicated by the anharmonic interactions in $H_\text{SB}^\text{SC}$.} Thus a particularly useful alternative perspective could be provided by semiclassical approaches based on approximate equations of motion, with ``equilibrium'' states identified by their stationarity under time evolution.


Finally, we note an alternative perspective on these results. Specifically, uncertainty relations are a cornerstone in quantum physics and are formally expressed in terms of the intrinsic uncertainty of quantum states \cite{Bal70}. The relevant uncertainty relations here result from the fact that the energy exchange between the system and the bath is mediated by the system-bath term $\hat H_\mathrm{SB} = \sum_k \hat{\mathcal{B}}_k \otimes \hat{\mathcal{S}}_k$; therefore, the system-bath energy fluctuations will be determined by $\Delta \hat H_\mathrm{S} \Delta \hat H_\mathrm{SB}\ge \frac{1}{2} \left | \langle [\hat H_\mathrm{S} , \hat H_\mathrm{SB}] \rangle \right| $ with $|\cdot|$ denoting the absolute value, $\langle \cdot \rangle = \mathrm{Tr} (\cdot \, \hat{\rho}_e)$ red{}{with $\hat\rho_e$ being the equilibrium distribution in Eq.~(\ref{rho_e_def}),} and $\Delta (\cdot) = \sqrt{ \langle( \cdot)^2 \rangle -\langle( \cdot) \rangle^2}$.

Recently, at equilibrium, these system-bath energy-uncertainty-relations were shown to be influenced by the spectral density of the bath and its thermal fluctuations.
For a large class of quantum systems\cite{PT&19}, it was shown that its classical counterpart is devoid of equilibrium stationary coherences and therefore, the energy-uncertainty-relation is zero.
This means that for those systems, uncertainty is solely of a classical nature (classical thermal fluctuations).

In our case, up to second order [see Eq.~(\ref{sigma_e_second_order})],
\begin{align}
\label{equ:UnctntyLwrBnd}
\Delta& \hat H_\mathrm{S} \Delta \hat H_\mathrm{SB} \ge
\frac{1}{2} \left | \sum_k \langle [\hat H_\mathrm{S} , \hat{\mathcal{S}}_k] \rangle \right|
	\\
	&\ge \frac{1}{2} \left | \sum_{k,\kappa,\mu,\nu} 	\left(H_\mathrm{S}^{\nu \kappa}  \mathcal{S}_k^{\kappa \mu} - \mathcal{S}_k^{\nu \kappa} H_\mathrm{S}^{\kappa \mu} \right) C_{\mu\nu}^\mathrm{Q}\right|,
\end{align}
with $\hat H_\mathrm{S}  = \sum_{\gamma, \gamma'}H_\mathrm{S}^{\gamma \gamma'} \proj{\eigenmarker \gamma}{\eigenmarker \gamma'} $ and $\hat{\mathcal{S}}_k  = \sum_{\kappa, \kappa'}\mathcal{S}^{\kappa \kappa'}_k
\proj{\eigenmarker \kappa}{\eigenmarker \kappa'} $.
The appearance of $C_{\mu\nu}^\mathrm{Q}$ shows the intimate role of coherences in
providing a measure of quantum effects between system and bath.

The classical vanishing lower-bound uncertainty is reached in Eq.~(\ref{equ:UnctntyLwrBnd})
when either $[\hat H_\mathrm{S},\hat{\mathcal{S}}_k] = 0,$ for all $k$ or when
$C_{\mu\nu}^\mathrm{Q}$, and therefore $\hat \sigma^{(2)}_e$ in Eq.~(\ref{equ:rho_quantum_second_order}),
is diagonal.
The uncertainty relation in Eq.~(\ref{equ:UnctntyLwrBnd}) is clearly a state
dependent quantity,
$ \sum_k \langle [\hat H_\mathrm{S} , \hat{\mathcal{S}}_k] \rangle_{\hat \rho} = \sum_k \mathrm{Tr} \left( [\hat H_\mathrm{S} , \hat{\mathcal{S}}_k] \hat \rho \right)$; therefore, the classical and semiclassical phase-space
distributions can be directly utilized in Eq.~(\ref{equ:UnctntyLwrBnd}).  
Specifically, the $\hbar^n$ expansion developed for $C_{\mu\nu}^\text{Q}$ in Section \ref{sec_hbar_expansion} can be used to smoothly interpolate from the full quantum result to the classical limit $C_{\mu\nu}^\text{Q} \to C_{\mu\nu}^\text{C} = 0$ as $\beta \to 0$.] The
latter then gives the classical only uncertainty of  $\Delta \hat H_\mathrm{S} \Delta \hat H_\mathrm{SB}=0$.

Equation~(\ref{equ:UnctntyLwrBnd}) is a key result, allowing the calculation of the
energy uncertainty bound from different levels of approximation to quantum coherences, e.g. the
semiclassical results given above.

\section{Conclusions}
\label{sec_conclusions}
In conclusion, existing classical descriptions fail to capture equilibrium stationary coherence effects in the MLSB model.
Quantization of the system energy through a semiclassical framework partially corrects this failure, giving an accurate description of quantum stationary coherence elements at high temperatures.
Even these models, however, fail qualitatively for the important case of symmetric system-bath interactions, i.e., for MLSB systems in which all sites couple to the bath with the same reorganization energy.
Thus, equilibrium coherence in the MLSB model appears to be an exclusively quantum-mechanical feature in systems relevant to light harvesting.
In addition, a Wigner-space expansion perturbative in $\hbar$ (rather than in either site-to-site or system-bath interaction strength) is introduced and appears promising for the efficient calculation of stationary coherence elements beyond the weak-coupling limit. Even at third order in $\hbar$, the results compare favorably at biologically-relevant temperatures to fully quantum-mechanical results for the dimer model studied.

\begin{acknowledgments}
MR thanks the Natural Sciences and Engineering Research Council of Canada (NSERC) for a Banting Postdoctoral Fellowship.
This work was partially supported by the AFOSR under contract number FA9550-17-1-0310 and by \emph{Comit\'e para el Desarrollo de la Investigaci\'on} --CODI-- of Universidad de Antioquia, Colombia under the grant number 2015-7631 and by the \emph{Departamento Administrativo de Ciencia, Tecnolog\'ia e Innovaci\'on}
--COLCIENCIAS-- of Colombia under the grant number 111556934912.
\end{acknowledgments}

\appendix

\section{Wigner Space Expansion}
\label{quantum_perturbation_expansion_appendix}

To obtain the $\hbar^3$ expansion of Section \ref{sec_quantumness}, perform the partial Wigner transformation\cite{Hillery1984} over the bath coordinates so that the reduced system density matrix $\hat \sigma_e$ takes the form
red{}{\begin{align}
    \label{sigma_e_wigner}
    \hat \sigma_e &= \frac{\hat{\mathcal P}_e \sigx \hat{\mathcal P}_e }{\mathrm{Tr}_\mathrm{S}\left \{ \hat{\mathcal P}_e \sigx \hat{\mathcal P}_e \right \}}
\end{align}}
with
\begin{align}
    \sigx &= \int \left( \prod_k \mathrm{d}Q_k \mathrm{d} P_k \right) \left [\mathrm{e}^{-\beta \left( \hat{H}_e
    + \hat{H}_\mathrm{B} + \hat{H}_\mathrm{SB} \right)} \right]_\mathrm{W},
\end{align}
where for any operator $\hat A$
\begin{align}
    \left [ \hat A \right]_\mathrm{W} &=
    \int \left( \prod_k \mathrm{d} \Xi_k \right) \mathrm{e}^{\frac{\mathrm{i} \boldsymbol{P} \cdot {\boldsymbol \Xi} }{\hbar}}
    \braAket{\boldsymbol{Q} - \frac{{\boldsymbol \Xi}}{2}}{\hat A}{\boldsymbol{Q} + \frac{{\boldsymbol \Xi}}{2}} .
\end{align}
The partial Wigner transform $\left [ \hat A \right]_\mathrm{W}$ is thus an operator over the system Hilbert space but a function of the bath coordinates.
Wigner transformations can be calculated from the rules\cite{Hillery1984}
\begin{align}
    \label{wigner_Q_transform}
    \left [ \hat Q_k \right]_\mathrm{W} &= Q_k
   \\
    \label{wigner_P_transform}
    \left [ \hat P_k \right]_\mathrm{W} &= P_k
\end{align}
and
\begin{align}
    \label{wigner_product_rule}
    \left [ \hat A \hat B \right]_\mathrm{W} = \left [ \hat A \right ]_\mathrm{W}
    \mathrm{e}^{\frac{\hbar \hat \Lambda}{2\mathrm{i}}} \left [ \hat B \right]_\mathrm{W},
\end{align}
where
\begin{align}
    \label{Lambda}
    \hat \Lambda &= \sum_k \left( \frac{\overleftarrow \partial }{\partial P_k}
    \frac{\overrightarrow \partial }{\partial Q_k} - \frac{\overleftarrow \partial }{\partial Q_k }
    \frac{\overrightarrow \partial }{\partial P_k} \right) .
\end{align}
In this last expression, the arrows indicate whether the derivative is applied to the quantity on the right or left of the expression.

Using the Zassenhaus expansion, we have
\begin{align}
    \label{zassenhaus_expansion}
   \mathrm{e}^{-\beta \left( \hat{H}_e + \hat{H}_\mathrm{B} + \hat{H}_\mathrm{SB} \right) }
   &=\mathrm{e}^{-\beta \hat{H}_\mathrm{B} }\mathrm{e}^{-\beta\hat{H}_e }\mathrm{e}^{-\beta\hat{H}_\mathrm{SB} }
\nonumber \\
    &\quad\quad \quad\quad \times\mathrm{e}^{ -\frac{\beta^2}{2} \hat C_1 }\mathrm{e}^{-\frac{\beta^3}{6} \hat C_{2} }...
\end{align}
where
\begin{align}
    \hat C_1 &= \left[ \hat{H}_e + \hat{H}_\mathrm{B}, \hat{H}_\mathrm{SB} \right] ,
\\
    \hat C_2 &= \left [ \hat{H}_e + \hat{H}_\mathrm{B}, \hat C_1 \right ] + 2 \left [ \hat{H}_\mathrm{SB}, \hat C_1 \right ] ,
\end{align}
and each remaining factor is an exponential in a sum $\hat C_n$ of $n$ nested commutators between $\hat{H}_e + \hat{H}_\mathrm{B}$ and $\hat{H}_\mathrm{SB}$.
The quantity $\sigx$ can then be written
\begin{align}
    \label{sigx_product}
    \sigx &= \int \left( \prod_k \mathrm{d}Q_k \mathrm{d} P_k \right) \left [\mathrm{e}^{-\beta \hat{H}_\mathrm{B}} \right]_\mathrm{W}
\nonumber \\
    &\quad\quad \times \left [\mathrm{e}^{-\beta \hat{H}_e}\mathrm{e}^{-\beta \hat{H}_\mathrm{SB}}
    \mathrm{e}^{-\frac{\beta^2}{2} \hat C_1 }\mathrm{e}^{- \frac{\beta^3}{6} \hat C_2 } ...\right ]_\mathrm{W} .
\end{align}
The absence in Eq.~\eqref{sigx_product} of the exponential operator $\mathrm{e}^{\frac{\hbar \hat \Lambda}{2\mathrm{i}}}$ that normally appears in Wigner-space products [see Eq.~\eqref{wigner_product_rule}] is due to the integration over the bath phase space.\footnote{Note that for any functions $f$ and $g$ of the bath coordinates whose product decays to zero at infinity
\begin{align}
    \int \mathrm{d}Q_k \int \mathrm{d}P_k \frac{\partial f}{\partial P_k} \frac{\partial g}{\partial Q_k}
    &= \int \mathrm{d}Q_k \int \mathrm{d}P_k \frac{\partial g}{\partial P_k} \frac{\partial f}{\partial Q_k} ,
\end{align}
so that
$ \int \mathrm{d}Q dP f \hat \Lambda g = 0$. Since $f \hat \Lambda^n g$ can always be written as a sum of terms of the form $f_{n-1} \hat \Lambda g_{n-1}$, where $f_{n-1}$ and $g_{n-1}$ are $(n-1)$-order derivatives of $f$, and $g$, this implies further that $ \int \mathrm{d}Q dP f \hat \Lambda^n g = 0$.}

This expression now offers a concrete method for computing the reduced density matrix $\hat \sigma_e$ to any given order in $\hbar$.
Indeed, the Wigner transform of $\mathrm{e}^{-\beta \hat{H}_\mathrm{B}}$ over the bath is known analytically as
\begin{align}
    \label{rho_B_W}
    &\left [ \mathrm{e}^{-\beta \hat{H}_\mathrm{B}} \right ]_{\bar W} =
    \prod_k \text{sech}\left( \frac{\beta \hbar \Omega_k}{2} \right)
\nonumber \\
    &\quad\quad \times \exp \left [ - \tanh \left( \frac{\beta \hbar \Omega_k}{2}\right)
    \frac{ \Omega_k^2 Q_k^2 + P_k^2}{\hbar \Omega_k} \right] .
\end{align}
The second factor in Eq.~\eqref{sigx_product} is tedious but straightforward to evaluate
at any given order.
In fact, Eqs.~\eqref{wigner_Q_transform} - \eqref{wigner_product_rule} imply that the
matrix elements of $\left[ \hat{H}_e \right]_\mathrm{W} = \hat{H}_e$ and
\begin{align}
    \left [ \hat{H}_\mathrm{SB} \right]_\mathrm{W} &= \hbar \sum_{nk} \alpha_{nk} Q_k \ket{n}\bra{n}
\end{align}
are both proportional to $\hbar$, while matrix elements for $\hat{H}_\mathrm{B}$ are
independent of $\hbar$.
Due to the factor of $\hbar$ in the exponent of Eq.~\eqref{wigner_product_rule},
commutation with $\hat{H}_\mathrm{B}$ always produces at least one additional
factor of $\hbar$ in Wigner space since
\begin{align}
    \label{H_B_commutator}
    \left [ \hat A, \hat{H}_\mathrm{B} \right]_\mathrm{W} &=
    \left [ \hat A \right]_{W} \left(\mathrm{e}^{\frac{\hbar \hat \Lambda}{2\mathrm{i}}} -
    \mathrm{e}^{-\frac{\hbar \hat \Lambda}{2\mathrm{i}}} \right)
    \left [ \hat{H}_\mathrm{B} \right]_{W}
\end{align}
and since the lowest-order term in the difference
$\mathrm{e}^{\frac{\hbar \hat \Lambda}{2\mathrm{i}}} -\mathrm{e}^{-\frac{\hbar \hat \Lambda}{2\mathrm{i}}}$
is linear in $\hbar$.
As a result of this relation and the linear scaling of $\left [ \hat{H}_e \right]_W$ and
 $\left [ \hat{H}_\mathrm{SB} \right]_W$, matrix elements for each $n$-fold commutator
 $\hat C_n$ scale as $\hbar^{n+1}$ or higher.
The $\hbar^n$ approximation for
$\left[e^{-\beta \hat{H}_e}\mathrm{e}^{-\beta \hat{H}_\mathrm{SB}}\mathrm{e}^{-\frac{\beta^2}{2} \hat C_1} ... \right]_W$
is thus obtained by including commutators up to $\hat C_{n-1}$, expanding each
exponential, and retaining terms up to $\beta^n$.

As an explicit example, consider the $\hbar^3$ result:
\begin{widetext}
\begin{align}
   \mathrm{e}^{-\beta\hat{H}_e }\mathrm{e}^{-\beta\hat{H}_\mathrm{SB} } &
   \mathrm{e}^{ -\frac{\beta^2}{2} \hat C_1 }\mathrm{e}^{-\frac{\beta^3}{6} \hat C_{2} } ... 
    \approx 1 - \beta \left( \hat{H}_e + \hat{H}_\mathrm{SB} \right) + \frac{\beta^2}{2}
    \left( \hat{H}_e^2 + \hat{H}_\mathrm{SB}^2 + 2 \hat{H}_e \hat{H}_\mathrm{SB} - \hat C_1 \right)
\nonumber \\
    &- \frac{\beta^3}{6} \left( \hat{H}_e^3 + \hat{H}_\mathrm{SB}^3 + 3 \hat{H}_e
    \hat{H}_\mathrm{SB}^2 + 3 \hat{H}_e^2 \hat{H}_\mathrm{SB} - 3 \left( \hat{H}_e +
    \hat{H}_\mathrm{SB} \right) \hat C_1 + \hat C_2 \right) .
\end{align}
This expression can be simplified by noting that any terms of overall odd order in bath coordinates vanish upon integration in Eq.~\eqref{sigx_product}.
Noting that $\hat{H}_e$ and $\hat{H}_\mathrm{B}$ are even in bath coordinates,
while $\hat{H}_\mathrm{SB}$ is odd, a brief calculation reveals that to order $\hbar^3$
\begin{align}
    \sigx 
    &\approx \int \left( \prod_k \mathrm{d}Q_k \mathrm{d} P_k \right) \left [\mathrm{e}^{-\beta \hat{H}_\mathrm{B}} \right]_\mathrm{W}
    \left [ 1 - \beta \hat{H}_e + \frac{\beta^2}{2} \left[ \hat{H}_e^2 + \hat{H}_\mathrm{SB}^2 \right]_\mathrm{W} \right.
\nonumber \\
    & - \frac{\beta^3}{6} \left[ \hat{H}_e^3 + \hat{H}_\mathrm{SB} \hat{H}_e \hat{H}_\mathrm{SB} +
    \hat{H}_\mathrm{SB}^2 \hat{H}_e + \hat{H}_e \hat{H}_\mathrm{SB}^2 + \vphantom{\frac{\beta^3}{6}} \left. \hat{H}_\mathrm{SB} \hat{H}_\mathrm{B}
    \hat{H}_\mathrm{SB} + \hat{H}_\mathrm{SB}^2 \hat{H}_\mathrm{B}
    - 2 \hat{H}_\mathrm{B} \hat H_\mathrm{SB}^2 \right]_\mathrm{W} \right].
\end{align}
\end{widetext}
To order $\hbar^3$, the Wigner transformations of all terms in this expression may be approximated by simply making the replacements $\hat Q_k \to Q_k$ and $\hat P_k \to P_k$
in the quantum operators of Eqs.~\eqref{HSB_quantum}, \eqref{HB_quantum}, and \eqref{He_quantum}.\footnote{The individual Wigner transforms may contain corrections of higher order in $\hbar$, but these can be discarded to accuracy $\hbar^3$ in the final expression.}
The result is that
\begin{align}
    \label{rho_third_order}
    \sigx &\approx Z_\mathrm{B}^\mathrm{C} \left \langle 1 - \beta \hat{H}_e + \frac{\beta^2}{2} \left(\hat{H}_e^2
    + \left(\hat{H}_\mathrm{SB}^\mathrm{QC}\right)^2 + B_2 \right) \right.
\nonumber \\
    & \left . - \frac{\beta^3}{6} \left( \hat{H}_e^3 + \hat{H}_\mathrm{SB}^\mathrm{QC} \hat{H}_e
    \hat{H}_\mathrm{SB}^\mathrm{QC} \right. \right.
\nonumber \\
    & \quad\quad\quad + \left .\left. \left( \hat{H}_\mathrm{SB}^\mathrm{QC}\right)^2 \hat{H}_e +
    \hat{H}_e \left( \hat{H}_\mathrm{SB}^\mathrm{QC}\right)^2 \right) \right \rangle_\mathrm{B} ,
\end{align}
where
\begin{align}
    \left \langle ... \right \rangle_\mathrm{B} &=
    \frac{1}{Z_\mathrm{B}^\mathrm{C}} \int \left( \prod_k \mathrm{d}Q_k \mathrm{d} P_k \right) \mathrm{e}^{-\beta H_\mathrm{B}} \left( ... \right)
\end{align}
represents a {\it classical} average over bath coordinates,
\begin{align}
    \hat{H}_\mathrm{SB}^\mathrm{QC} &= \sum_{n,k} \alpha_{nk} Q_k \ket{n}\bra{n},
\end{align}
and the term
\begin{align}
    B_2 &= \sum_k \frac{\hbar^2 \Omega_k^2 }{4} \left[ \frac{\beta \left( \Omega_k^2 Q_k^2 + P_k^2\right)}{3} - 1 \right]
\end{align}
arises from the third-order expansion of Eq.~\eqref{rho_B_W}. Taking off-diagonal matrix elements of Eq.~\eqref{rho_third_order} in the system eigenbasis then yields Eq.~\eqref{C12_hbar3} of the main text. Note that, to third order, quantum corrections to the partition function $\mathrm{Tr}_\mathrm{S}\left \{ \sigx \right \}$ do not affect ${C}_{\mu\nu}^\mathrm{Q}$ since the lowest non-vanishing term is already of order $\hbar^2$ and since
\begin{align}
    \mathrm{Tr}_\mathrm{S}\left \{ \hat{H}_e \right \} &= \hbar \sum_{\eigenmarker \mu} \delta \eigenmarker \omega_\mu = 0
\end{align}
so that the first-order correction to the partition function vanishes.

\section{Phase-Space Representation of a Quantum Coherence}
\label{sec_rho_10_W}

The Wigner distribution for the quantum operator $\ket{0}\bra{0}$ of a harmonic oscillator of frequency $\omega$ is given by\cite{Hillery1984}
\begin{align}
    \left [ \ket{0}\bra{0} \right]_W &= \frac{1}{\pi\hbar} e^{-\frac{\omega^2 q^2 + p^2}{\hbar\omega}} .
\end{align}
The Wigner-space representation of the raising operator $\hat a^\dagger = \frac{\omega \hat q - i \hat p}{\sqrt{2\hbar\omega}}$ is simply
\begin{align}
    [ a^\dagger]_W = \frac{\omega q - i p}{\sqrt{2\hbar\omega}}.
\end{align}
Thus according to the Wigner-space product rule [Eq. \eqref{Lambda}] (with $Q_k \to q$)
\begin{align}
    \left [ \ket{1}{0}\right]_W &= \left [ \hat a^\dagger \ket{0}\bra{0}\right]_W\\
    &= \left [ \hat a^\dagger \right]_W \mathrm{e}^{\frac{\hbar}{2 \mathrm{i}} \left( \frac{\overleftarrow \partial }{\partial p}
    \frac{\overrightarrow \partial }{\partial q} - \frac{\overleftarrow \partial }{\partial q }
    \frac{\overrightarrow \partial }{\partial p} \right) } \left [\ket{0}\bra{0}\right]_W .
\end{align}
Expansion of the exponential and evaluation of the derivatives gives Eq.~\eqref{rho10_q}.

\bibliography{references}

\end{document}